\documentstyle[prl,aps,epsfig,floats]{revtex}

\begin{document}

\newcommand{\be}{\begin{equation}}
\newcommand{\ee}{\end{equation}}
\newcommand{\bea}{\begin{eqnarray}}
\newcommand{\eea}{\end{eqnarray}}
\newcommand{\ba}{\begin{array}}
\newcommand{\ea}{\end{array}}
\newcommand{\sprime}{^\prime}
\newcommand{\dprime}{^{\prime\prime}}
\newcommand{\tprime}{^{\prime\prime\prime}}

\twocolumn[
\hsize\textwidth\columnwidth\hsize
\csname@twocolumnfalse\endcsname
\begin{flushright}
	IFP-781-UNC, hep-ph/0002135 
\end{flushright}
\title{Seeking Gauge Bileptons in Linear Colliders}
\author{Paul H. Frampton and Andrija Ra\v{s}in}
\address{
	\tighten{\it Department of Physics and Astronomy\\
	University of North Carolina, Chapel Hill, NC 27599-3255 \\
	\today}
	}
\maketitle
\begin{abstract}
	A promising direction to find physics beyond the standard model
	is to look for violation of $L_{e,\mu,\tau}$ conservation. In
	particular the process $e^- e^- \rightarrow \mu^- \mu^-$ with
        the exchange of a gauge bilepton has a
	striking signal without background and is predicted in the most
        economical model to have a cross-section an order of magnitude
        higher than previous estimates.\\
	\today
\end{abstract}
\vspace{0.2in}
]\narrowtext

\section{Introduction}

Conservation laws, and their relation to symmetries, are 
fundamental to the theoretical physics of nuclei, condensed
matter as well as high energy physics. This has already led
to great advances and further progress will
certainly come when the correct symmetries are better
probed and identified.

It is expected that clear deviations from the Standard Model (SM)
of particle physics should soon show up in experiment. Already the
non-zero neutrino mass is one example of this - in the minimal
SM neutrinos have no mass. Beyond that the best hope is to
observe rare processes, especially those absolutely forbidden
in the minimal SM. The best suited are processes which violate
global conservation laws, since there is no reason to believe
in their exactness. For example, baryon number ($B$) might be violated
as in the proton decay predicted by {\it e.g.} grand unification.
Even more amenable to test may be the processes which violate
the separate lepton numbers $L_{e,\mu,\tau}$ since, even if
$L = L_e + L_\mu + L_\tau$ is good, there is little reason
to believe in $L_{e,\mu,\tau}$. Neutrino oscillations
exemplify this.

One outstanding question of the SM is the occurrence of replicas of
the quark-lepton families and one approach to that issue is the
cancellation of chiral anomalies between asymmetric
non-sequential families of the 331 model
\cite{framprl92,pisa92} (based on the gauge group
$SU(3)_c \times SU(3)_L \times U(1)$). 
Recall that anomaly cancellation is crucial in many situations
of model building beyond standard model, {\it e.g.} 
chiral color\cite{anomalych} and in string theory\cite{anomalyst}. 
The 331 model simultaneously
predicts that while $L$ is a good quantum number perturbatively,
processes do occur with $|\Delta L_i| = 2$ ($i=e,\mu,\tau$). There is no
SM background, and so a process like $e^- e^- \rightarrow \mu^- \mu^-$
shown in Figure 1
provides a way to test\cite{eeconf} whether bileptonic $Y^{--}$ gauge
bosons
are present, as they are if $SU(2)_L \subset SU(3)_L$ for the gauge 
group. Another model studied in this context is the SU(15) theory,
where there are double charged gauge bileptons too, however with 
a smaller coupling (see next Section). 
Among other theories which have doubly charged particles
with $L = 2$ are the left-right models based on the group 
$SU(2)_L \times SU(2)_R \times U(1)_{B-L}$\cite{ms79}. 
However, the couplings of these states to the SM fermions are of Yukawa
type and are not fixed. On the other hand, even though $M_R$, the scale of
$SU(2)_R$ symmetry breaking, may be high, there are light 
double charged scalars and fermions\cite{ams97,ot}, which may lie close to
the weak scale. The situation
also holds in the supersymmetric grand
unified versions of left-right models, where complete double charged
Higgs multiplets may be much lighter than the scales of left-right and
grand unified symmetries\cite{abmrs99}. The same experiments
that will search for gauge bileptons will set important limits for these
states as well\cite{gun98}.

\begin{figure}
\centering
\epsfig{file=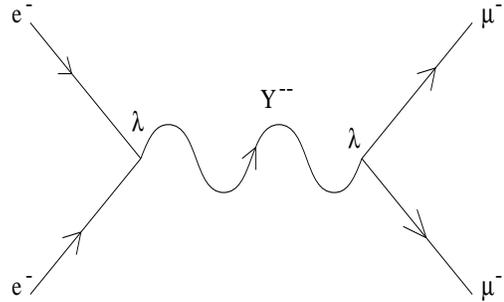,height=3.9cm,width=6.6cm}
\vspace{0.5cm}
\caption{Contribution of the gauge bilepton to the $e^- e^-
\rightarrow \mu^- \mu^-$ scattering.}
\end{figure}

In old work on the subject of gauge bileptons in the early 90's,
the $L_{e,\mu}$ violating processes were calculated in the context
of an SU(15) theory\cite{framng92,frammpl92}. Here we point out that in
the more economic and
attractive 331 model, the cross-section
$\sigma(e^- e^- \rightarrow \mu^- \mu^-)$
is larger by an order of magnitude than in SU(15). This result
holds true in any model that embeds the gauge group $SU(2)_L$
in a single $SU(3)_L$.

The lower limit on $M(Y^{--})$ is currently claimed 
to be at $850$ GeV\cite{will99} (however see discussion
below, also see \cite{plei99,das99}),
and with that value of $M(Y^{--})$ the cross-section for
$\sigma(e^- e^- \rightarrow \mu^- \mu^-)$ is already
$\sim 5$ fb at $\sqrt{s} = 100$ GeV and climbs very
fast with $s$ to several hundred fb at $\sqrt{s} = 500$ GeV. 
This implies a readily detectable event rate
for an $L_{e,\mu}$ violating process in a collider with
integrated luminosity of
$10 ({\rm fb})^{-1}$ or higher, 
{\it e.g.} 
$10^{33} /{\rm cm}^2 {\rm s}$ for a few months
or  
$10^{34} / {\rm cm}^2 {\rm s}$ for a week.

Some of the leading candidates presently discussed for the next generation
of particle colliders (beyond LEP, Tevatron and LHC) are the 
Next Linear Collider (NLC), the (circular) Muon Collider (MC)
and the Very Large Hadron Collider (VLHC). Of these the most advanced in
design and preparation is the NLC. This is most often assumed to
be an $e^+ e^-$ positron-electron collider with center of mass energy in
the range above LEP: $200 {\rm GeV} < \sqrt{s} < 2000 {\rm GeV}$, 
although also being discussed is the desirability to operate $e^+ e^-$
at $\sqrt{s} = M_Z$ with high luminosity.

The motivation for such an NLC is to investigate physics in a
presently-unexplored energy domain, although by the time of its start-up
initial discoveries can be expected to have been made by LHC whereupon the
NLC provides an excellent tool for detailed studies.

It is these machines that are ideally suited for finding the
bilepton even for $\sqrt{s} < 200 $ GeV (see Figure 2), but with
luminosity higher than LEP. 
These machines could therefore initially run in both charge
modes at such an energy.

\bigskip
\bigskip

\section{Cross Section}

The interaction Lagrangian between the bilepton $Y$ and 
leptons is given by 
\be
L_Y = \lambda Y^{++}_\mu e^T C \gamma^\mu P_{-} e +
+ \lambda^\dagger Y^{--}_\mu \overline{e} \gamma^\mu P_{-} C
\overline{e}^T
\ee
where $\lambda$ is the $3 \times 3$ coupling matrix. For
simplicity, in the
following we will assume $\lambda^{ij} \sim \lambda {\bf 1}$.

In the 331 model the group $SU(2)_L$ is completely
embedded in the $SU(3)_L$ so that the gauge bilepton
coupling is equal to the coupling of $SU(2)_L$
\be
\lambda = { g \over \sqrt{2} } \approx { {2.08 e} \over \sqrt{2} }
\ee

This coupling is larger than the coupling arising from the
SU(15) models where $\lambda \approx 1.19 e / \sqrt{2}$ simply because
the $SU(2)_L$ in that case is not residing entirely in 
a single $SU(3)_L$ but also partially in the
$SU(6)_L$ subgroup of $SU(15)$\cite{framlee90}.

The differential cross section (see also \cite{framng92,cuyr97})
for the
gauge bilepton exchange in the process 
$e^- e^- \rightarrow \mu^- \mu ^-$ is given by
\be 
{ {d \sigma} \over {d \cos \theta} } = 
{{\lambda^4} \over {32 \pi} }
\,\, 
( 1 + \cos^2 \theta )
\,\,
{ s \over {(s-m_Y^2)^2 + m_Y^2 \Gamma_Y^2} }
\ee
where $\theta$ goes from $0$ to $\pi/2$ reflecting the identical
particles in the final states. 
Here $m_Y$ and $\Gamma_Y$ are
the mass and decay width of the bilepton and
$s$ is the square of the center of mass 
energy.  This leads to the total cross section
\be
\sigma = {{\lambda^4} \over {24 \pi}} 
	{ {s} \over {(s-m^2_Y)^2 + m^2_Y \Gamma^2_Y}}
\label{cs}
\ee
When the flavor and mass eigenbasis of leptons
do not match, the results above must be corrected by
$\lambda^2 \rightarrow \lambda^2 V_{ej} V_{je}^* V_{\mu j} V_{j \mu}^*$.

The total cross section as a function of the bilepton
masses is plotted in Figure 2 for various values
of $\sqrt{s}$ (We estimated the decay width of the bilepton
as\cite{cuyr97} $\Gamma_Y = \lambda^2 m_Y / 8 \pi$.) 

\begin{figure}
\epsfig{file=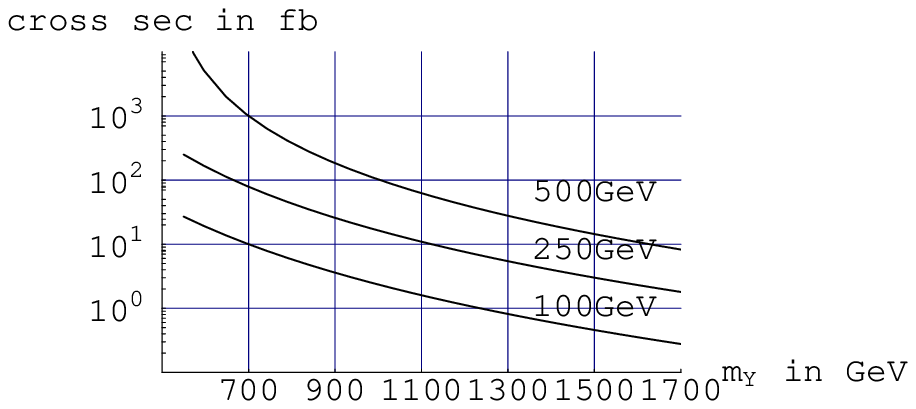,height=7.8cm,width=9.0cm}
\caption{Total cross section in the 331 model for the $e^-e^- \rightarrow
\mu^-\mu^-$ process as a function of bilepton mass at $\sqrt{s} =
100,250,500$ GeV.}
\end{figure}

We see that already at $\sqrt{s} = 100$ GeV we get a cross section
of about $\sigma \sim 5$ fb for a bilepton mass $m_Y = 850$ GeV,
and for the same $m_Y$ becomes $250$ fb at $\sqrt{s} = 500$ GeV.
This is about an {\it order of magnitude} higher than in the case of
SU(15) model.

Such a large cross section makes such a process highly
visible. The situation is especially good for
NLC-turned-into-$e^-e^-$-machine,
 even if it would run only at
$\sqrt{s} = 100$ GeV, but with the desired luminosity of
$10^{33}/cm^2s$. 

This motivates a first run of future linear
colliders (electron or muon) machines as 
{\it same} charge lepton collider. There are another
two reasons for considering doing this: a) the additional
cost to switch a machine from $e^+e^-$ to $e^-e^-$ is
relatively negligible  ;
b) the highest energy that previously probed $e^- e^-$
collisions was in 1971 at $\sqrt{s} = 1.12$ GeV
\cite{barb71}, so this would present an opportunity
to improve bounds on the bilepton mass considerably.

The spin 1 of the state $Y$ can be checked by the angular 
distribution of Eq. (3). For spin 0 the distribution is 
isotropic.

The inverse process $\mu^+ \mu^+ \rightarrow e^+ e^+$
($\mu^+$ is easier to ``cool" than $\mu^-$) 
has the same cross section and could be probed at
muon colliders, which are recently discussed as the
next colliders in addition to $e^+ e^-$ machines.

\section{Backgrounds}

Since the scattering violates $L_e$ and $L_\mu$
it does not have any background in the standard
model. However, a non-zero neutrino mass violates lepton
number and one can easily find  processes which involve
neutrino mass that produce $e^- e^- \rightarrow \mu^- \mu^-$,
as for example shown in Figure 3. Recently, first 
indications of neutrino mass have been found, and one
can ask the question how serious a background the
neutrino exchange can be to the bilepton exchange, given
the mass ranges quoted in the atmospheric and solar neutrino
data. In fact such calculations have been done 
recently\cite{bela95} which find that the cross section
is very small. The strongest bound for Figure 3a)
comes from the neutrinoless double beta decay where it was found that
the cross section for the process $ e^- e^- \rightarrow W W$
is smaller than $2.5 \times 10^{-7}$ fb for
$\sqrt{s} = 100$ GeV and is thus quite negligible
compared to the vector bilepton exchange. Bounds on 
contributions in Figure 3b) are found in \cite{bela95}
to be of a similar order, either because of a small
light neutrino mass or a small mixing with the heavy
Majorana neutrino masses. On the other hand, using 
fine tuning, very light right-handed neutrinos,
or contributions from extra doubly
charged scalar states, it is possible\cite{bela95} 
to make the cross sections somewhat larger (typically
$10^{-6}-10^{-5}$ fb), but these are still much
smaller than the bilepton signal. Also, we note
that a cut would have to be applied for 
back-to-back muons in the final state, 
and therefore the signal to background ratio
would be even higher.

In any case, non-back-to-back $\mu^-\mu^-$ in the
final state are more likely to be from 
$e^- e^- \rightarrow Y^{--} \rightarrow \tau^- \tau^-
\rightarrow \mu^- \mu^- + {\rm neutrinos}$,
which has the same cross-section as in Eq.(\ref{cs}),
giving another positive signal for the $Y^{--}$.

\begin{figure}[t]
\centering
\epsfig{file=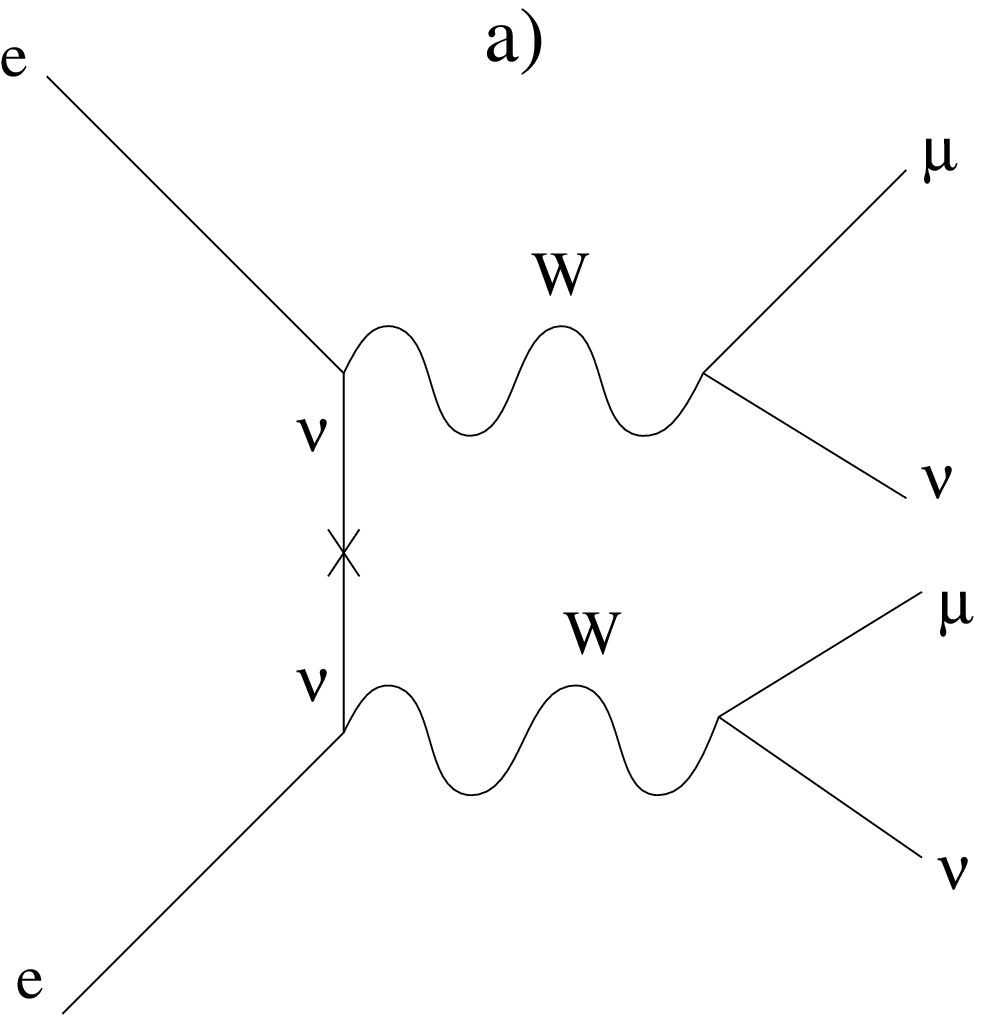,height=4.4cm,width=5.8cm}
\centering
\epsfig{file=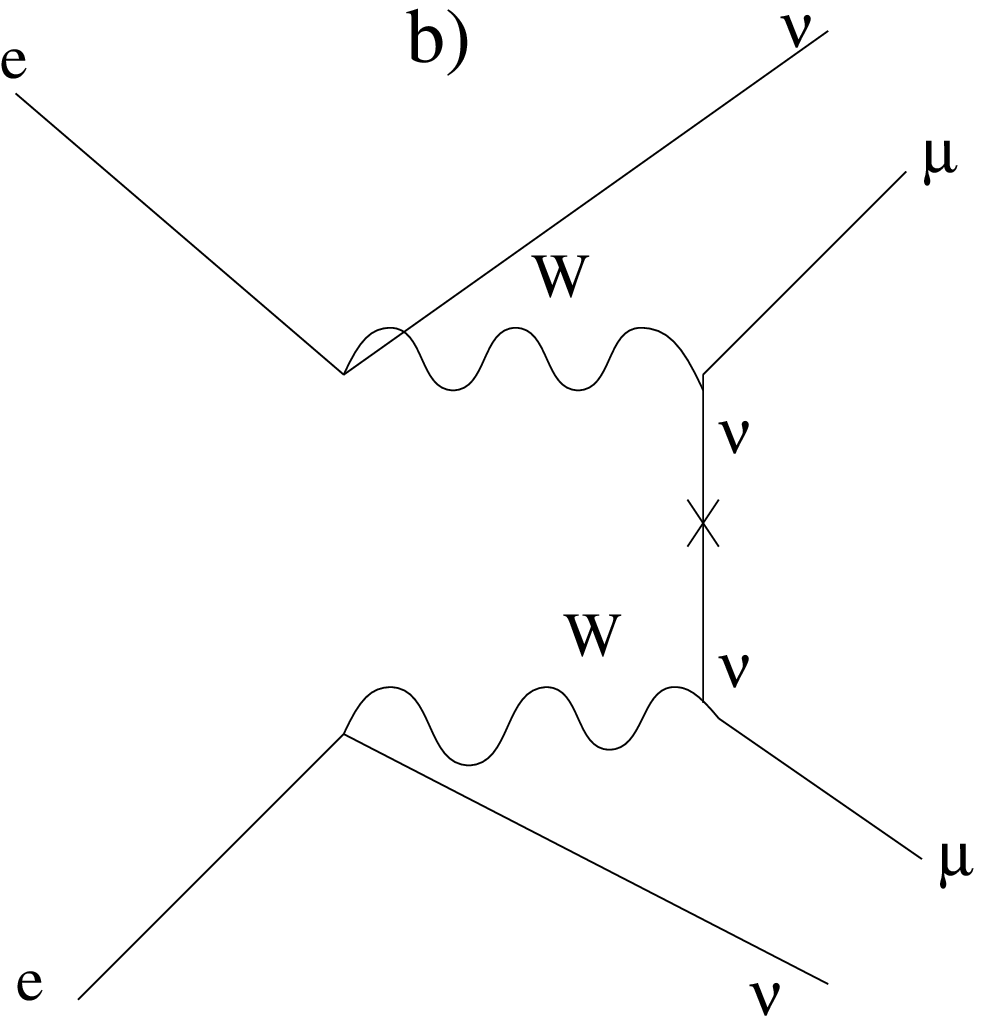,height=4.4cm,width=5.8cm}
\caption{Contributions from non-zero neutrino mass to background of 
$e^- e^- \rightarrow \mu^- \mu^-$.}
\end{figure}

A diagram not discussed in previous papers that contributes
to the $e^-e^-$ scattering and involves a pair of
back-to-back scattered muons in the final states 
is shown in Figure 4. This process is
not related directly to the neutrinoless double beta decay data.
However, it is bound to be still smaller than the
processes in Figure 3, because of the small
neutrino masses. We estimate
\be
	T_{\rm Fig. 4} 
\sim 
	{1 \over {16 \pi^2}} 
	|V_{ei}|^2 |V_{\mu j}|^2
	{ 
		{m_{\nu_i} m_{\nu_j}} 
	\over 
		{M^4_W} 
	}
\ee
where $V_{li}$ are the Kobayashi-Maskawa like mixing angles in
the lepton sector. Even if we take typical neutrino masses to
be $\sim 1eV$ this process is {\it completely} negligible 
(down by over ten orders of magnitude) compared
to the bilepton diagram (with $m_Y$ in the TeV region) which has 
$T \sim  V_{ei} V^*_{\mu j} / m_Y^2$.

\begin{figure}
\centering
\epsfig{file=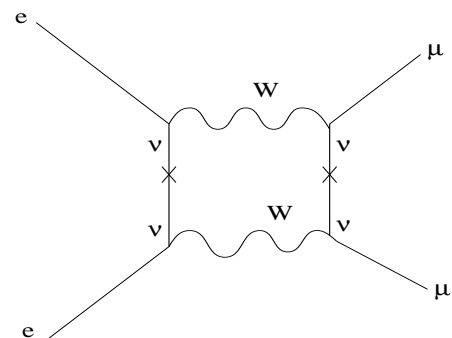,height=4.4cm,width=5.8cm}
\caption{Another contribution to background of $e^- e^- \rightarrow \mu^-
\mu^-$.}
\end{figure}

\section{Discussion}

In this letter we have shown that the next generation of 
linear colliders, when run in the same charge mode, are
especially suitable for search of doubly charged gauge bosons,
that naturally appear in theories where $SU(2)_L$ is
completely embedded in a new $SU(3)_L$ gauge group. Of special interest
is the process $e^- e^- \rightarrow \mu^- \mu^-$ which
has no background in the standard model. Even if
neutrino masses are taken to be non-zero, with the current limits,
the background to the bilepton exchange is completely
negligible for $\sqrt{s} \leq 1$ TeV. 
This, in our opinion, strongly motivates renewed
interest in running the future colliders first as $e^-e^-$
or $\mu^-\mu^-$ machines.

One aspect of the NLC, not usually mentioned in
a physics paper but relevant to the real world and to our 
discussion, is its cost and the concomitant likelihood of 
being funded and constructed. For simplicity take the cost 
of an NLC to be linear in energy and assume it as 
one penny per eV. This estimate is very
approximate, but, in any case, the minimization of
$E$  will clearly minimize the start-up cost of a machine
which could then be extended to achieve higher $E$. The cost
of running $e^- e^-$ is perhaps slightly less than 
for $e^+ e^-$, but both are needed for their physics interests. 
The $e^+e^-$ mode at $\sqrt{s} = 100$ GeV provides useful machine physics
and exploration of the $Z$ pole at a higher luminosity than that of
LEP ($5 \times 10^{31} /{\rm cm}^2 {\rm s}$).

Finally, we comment on the current experimental limits on the bilepton
mass. One of the strongest limits comes from the fermion pair
production and lepton-flavor violating charged lepton decays
\cite{tull99} and is about 750 GeV. We see from Figure 2, that
even for $\sqrt{s} = 100$ GeV we get huge signals for masses
above the current limit and up to 1 TeV and well beyond.
On the other hand an even stronger limit is claimed to come
from the muonium-antimuonium 
conversion\cite{will99} which currently set the limit at about
$850$ GeV for the SU(15) like coupling. Since the diagram 
contributing to that process is exactly the same as the one
contributing to the $e^-e^- \rightarrow \mu^-\mu^-$ scattering,
one would at first sight think that the same limit applies, 
and so for a 331 coupling (which is larger than the SU(15))
the above limit would translate to an even stronger bound. However,
there are several reasons why the bound from the muonium-antimuonium
conversion does not have to apply in the same way. First, it was
noted by Pleitez\cite{plei99} that the limit of 850 GeV may be much less
severe
if there are additional diagrams with, for example, scalars 
contributing to the same process, that may cancel each other.
In fact such extra scalars do appear in the minimal 331
model\cite{framprl92}. 
Second, the muonium-antimuonium conversion happens at energies
which are much smaller than $m_Y$, typically less than 1 GeV.
Then, however, for the $e^-e^- \rightarrow \mu^-\mu^-$ process
which is happening at much higher energies (and with a nonnegligible
center of energy $s$) the cancellation of diagrams may disappear
and the process will be enhanced relatively to the muonium-
antimuonium conversion. 

Nevertheless, as we have shown, even for gauge bilepton masses in
the TeV region, the signatures in $e^-e^- \rightarrow \mu^-
\mu^-$ scattering would be significant at attainable design luminosities.
In our opinion this warrants a serious examination of the 
possibility of including this mode in the designs of the future generation
of linear colliders. It would provide a potentially very exciting
experiment in the early stages of operation of such machines.

\bigskip
\bigskip
\bigskip
\bigskip
\bigskip

{\bf Acknowledgments  } \hspace{0.5cm} 

We thank David Burke for answering experimental queries
and Ryan Rohm for a very useful discussion. 
This work was supported in part by the US Department of Energy
under the Grant No. DE-FG02-97ER-41036.

\end{document}